# Monitoring method for neutron flux for a spallation target in an accelerator driven sub-critical system

ZHAO Qiang (赵强)[1,2], HE Zhiyong (贺智勇)[1,\*], YANG Lei (杨磊)[1], ZHANG Xueying (张雪荧)[1],

CUI Wenjuan (崔文娟)[1], CHEN Zhiqiang (陈志强)[1], and XU Hushan (徐瑚珊)[1]

[1] Institute of Modern Physics, Chinese Academy of Sciences, Lanzhou 730000, China

[2] University of Chinese Academy of Sciences, Beijing 100049, China

\*corresponding author, zyhe@impcas.ac.cn

**Abstract**: In this paper, we study a monitoring method for neutron flux for the spallation target used in an accelerator driven sub-critical (ADS) system, where a spallation target located vertically at the centre of a sub-critical core is bombarded vertically by high-energy protons from an accelerator. First, by considering the characteristics in the spatial variation of neutron flux from the spallation target, we propose a multi-point measurement technique, i.e. the spallation neutron flux should be measured at multiple vertical locations. To explain why the flux should be measured at multiple locations, we have studied neutron production from a tungsten target bombarded by a 250 MeV-proton beam with Geant4-based Monte Carlo simulations. The simulation results indicate that the neutron flux at the central location is up to three orders of magnitude higher than the flux at lower locations. Secondly, we have developed an effective technique in order to measure the spallation neutron flux with a fission chamber (FC), by establishing the relation between the fission rate measured by FC and the spallation neutron flux. Since this relation is linear for a FC, a constant calibration factor is used to derive the neutron flux from the measured fission rate. This calibration factor can be extracted from the energy spectra of spallation neutrons. Finally, we have evaluated the proposed calibration method for a FC in the environment of an ADS system. The results indicate that the proposed method functions very well.

## 1. Introduction

In an accelerator driven sub-critical (ADS) system, a heavy metal spallation target located at the centre of a sub-critical core is bombarded by high-energy protons from an accelerator. The spallation neutron from the target is used as an intense external neutron source to drive the sub-critical reactor. In order to successfully couple the three components in the ADS system, the accelerator, the target and the reactor, real-time measurement of neutron flux from the spallation target and the reactor core is very useful. It is also necessary for the commissioning measurements of the proton beams from the accelerator, for the routine verification of control rod positions, and for the calibration of the excore power range nuclear instruments.

In this paper, we focus on the monitoring method of neutron flux for the spallation target used in an ADS systems.

In a commercial reactor such as a pressurized water reactor used in nuclear industry, fission chamber (FC) detectors are used to measure the incore flux profile of thermal neutrons within the reactor core. To make a FC sensitive to thermal neutrons, a coating with the uranium isotope $^{235}$U is added in the chamber. However, the neutrons from a heavy-metal spallation target bombarded by high-energy protons are fast neutrons with energy more than 0.1 MeV. Since the uranium isotope $^{235}$U has a much lower fission cross section for fast neutrons than that for the thermal neutrons of 0.025 eV, a natural question is whether a commercial fission chamber for thermal neutrons can be used to measure the fast neutrons from the spallation process?

In this paper, we study the application of FCs in neutron monitoring for a spallation target. In order to accurately measure the neutron flux, we develop a calibration method which is based on the energy spectra of spallation neutrons. Furthermore, we propose that FCs should be put at different vertical locations to measure the spallation neutrons, by considering the characteristics in the spatial variation of neutron flux from the spallation target. The paper is organized as follows. First, a multi-point measurement technique for spallation neutrons is proposed in Section 2. To explain why it is important to take the multi-point measurement, the Monte Carlo simulation for spallation reactions is discussed in Section 3. Then, the calibration method for a FC detector in the neutron monitoring of a spallation target is developed in Section 4. In Section 5 the calibration method is evaluated in the environment of an ADS system, where the neutron flux measured by a FC within a reactor comes from both spallation neutrons and fission neutrons. Finally, the main contributions of this paper are summarized in section 6.

## 2. Proposed monitoring method

To investigate the monitoring technique for neutrons for a spallation target, we consider the China Initiative Accelerator Driven System (CIADS) which should be able to demonstrate the ADS concept at 10 MW power level with a maximum incore neutron flux of $2 \times 10^{14}$ n/cm$^2$/s. Extension of the proposed technique to an ADS system with arbitrary power level is straightforward.

### 2.1 Multi-point measurement technique

In the CIADS facility, a tungsten target located vertically at the centre of a sub-critical core is bombarded vertically by protons with energy of 250 MeV and a beam current of 10 mA. By considering the characteristics of the spatial variation of neutron flux from the spallation target, we propose to measure the spallation neutron flux at multiple vertical locations. The detectors for the measurements of neutron flux may either be left in a fixed location, the so-called fixed measurement method, or provided with a motorized drive to allow vertical movement within the reactor core, the so-called mobile measurement method. Compared to the mobile method, the fixed method can provide full-time monitoring and improve the reactor operating margin. However, more detectors are required with the fixed measurement method.

To further explain the proposed multi-point measurement technique, Fig. 1 shows a three-point measurement system, where the neutron flux is measured at three vertical locations, i.e. the upper, central and lower locations.

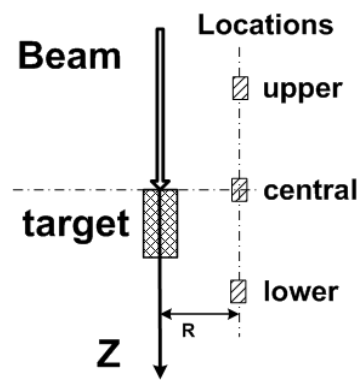

Fig. 1. The neutron monitoring method for the spallation target in an ADS system where a spallation target located vertically at the centre of the core is bombarded vertically by a proton beam. The vertical coordinate is taken as the z-axis and the top surface of the target is taken as the coordinate origin.

### 2.2 Evaluation of the proposed monitoring method

To explain why it is important to take the multi-point measurement for spallation neutrons, we have performed the simulation of spallation reactions with the GEANT4 (GEometry ANd Tracking) toolkit [1-2]. The Geant4 toolkit has been employed widely in basic physics research to simulate the propagation of particles and nuclei in extended media. Monte Carlo simulation based on the Geant4 toolkit has also been performed for the ADS system (e.g. [3-4]). The Geant4 toolkit provides several sets of physics models which simulate the interactions of protons and neutrons with atomic nuclei. In our simulation, the GEANT4 toolkit is used to calculate neutron yield for a 250-MeV proton beam impinging on tungsten targets. The simulation of spallation reactions with the GEANT4 toolkit for a 250-MeV proton beam have been validated by the experimental results. For example, the calculation results of the neutron yield for a 256-MeV proton beam impinging on a lead target have been verified by Forger et al [5]. At four angles of 7.5°, 30°, 60° and 150°, the calculation results of the double differential neutron yield are in very good agreement with the experimental results in [6].

To evaluate the proposed multi-point measurement technique, we have performed the simulation with GEANT4 by considering a proton beam with energy of 250 MeV and current of 10 mA bombarding a tungsten target. The target is made of natural tungsten corresponding to G4_W material from the list of NIST materials in Geant4. The target has a cylindrical shape with radius of 10 cm and lengths of 1, 3, 5, 7, 10, 15, 20, 30, 40, 50, 70 or 100 cm. As shown in Fig. 1, the vertical coordinate is taken as the z-axis and the top surface of the target is taken as the coordinate origin. The distance $R$ between the detector and the z-axis is 20 cm. Although there may be some containers for the detector and the target, we do not consider any neutron absorption of these container materials. The simulation results are discussed in the next section.

## 3. Characteristics of spallation neutrons

Spallation reactions, which produce spallation neutrons, are generally described by a two-stage mechanism. In the first stage, successive hard collisions between the incident particle and the individual nucleons of the target nucleus, i.e. nucleon-nucleon collisions, lead to the emission of a few fast nucleons. The high energy part of the neutron spectra comes from this stage. In the second stage, statistical de-excitation of the excited nuclei, namely, evaporation, multi-fragmentation and fission, leads to the emission of low energy particles or heavy nuclei. Low energy neutrons, which are the majority of neutrons produced in spallation reactions, are emitted during the evaporation process. The first stage is generally described by intranuclear cascade models while evaporation-fission models are used for the second stage. Sometimes one adds a pre-equilibrium stage between intranuclear cascade and de-excitation.

### 3.1 Neutron energy spectrum at different locations

Fig. 2 shows the GEANT4 simulation results for neutron yields per solid angle per second per MeV as a function of the energy of neutrons at 4 locations, i.e. $z = -30, 0, 30$, and 60 cm, respectively, for neutron emission from a tungsten target of 30 cm length when bombarded with 250 MeV protons. As shown in Fig. 2, low-energy neutrons with energy less than 10 MeV can be seen in all curves. The neutron yields per solid angle in the low-energy part of the neutron spectrum at $z = -30, 0$ and 30 cm are almost same, because these neutrons mostly come from the evaporation process. But the neutron yield at $z = 60$ cm is much lower, due to the neutron absorption of the target material. The high-energy neutrons from the process of nucleon-nucleon collisions can be seen in the curves with $z = 30$ and 60 cm. As shown in Fig. 2, the spallation neutrons from the target are fast neutrons with energy of more than 0.1 MeV. Thus, fast neutron detectors are expected to measure the neutrons from the spallation target.

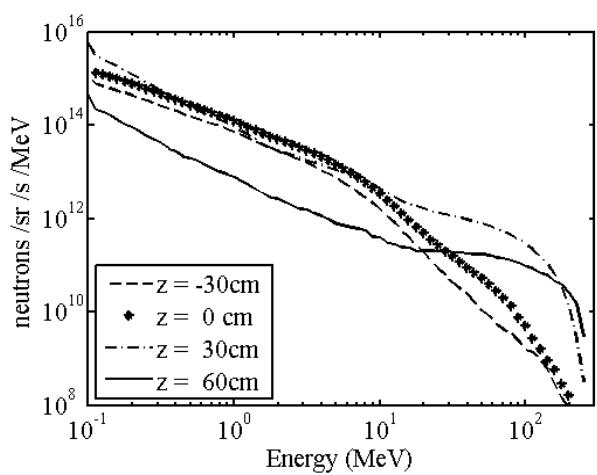

Fig. 2. Neutron yields per solid angle per MeV per second for tungsten target with a length of 30 cm. The 250-MeV proton beam with current of 10 mA has been simulated with Geant4 code.

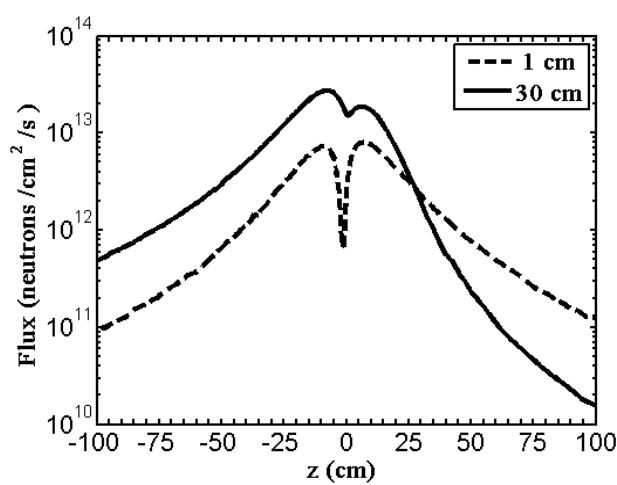

Fig. 3. Neutron flux distributions at different locations $Z$ for 250-MeV proton beam with current of 10 mA impinging on a tungsten target. The simulation results based on Geant4 code with two target lengths, 1 cm and 30 cm, are compared.

## 3.2 Neutron flux at different locations

Fig. 3 shows the GEANT4 simulation results for neutron yields per cm$^2$ per second as a function of the vertical location $z$ for a 250-MeV proton beam with a current of 10 mA impinging on a lead target. As shown by the dashed line in Fig. 3, the neutron flux distribution is symmetrical around $z = 0$ when the target length is 1 cm. The neutron flux decreases with increasing $z$ value, because the distance between the detector and the target increases with increasing $z$ value. When the target length increases to 30 cm, as shown by the solid line in Fig. 3, the neutron flux distribution decreases very steeply with increasing $z$ value from $z = 20$ to $z = 100$ cm, because of the neutron absorption of the target material.

The difference of neutron flux between the central location, i.e. $z = 0$ cm, and the lower location, e.g. i.e. $z = 100$ cm, reaches up to three orders of magnitude for a 250-MeV proton beam on a lead or bismuth target. The neutron flux at the central location reaches $2 \times 10^{13}$ n/cm$^2$/s with the beam current of 10 mA. Given the design flux of $2 \times 10^{14}$ n/cm$^2$/s for the CIADS facility, it should be fast to reach the design flux at the central location. On the other hand, it will take more time to reach the design flux at the lower or upper locations which are far from the target.

## 4. Monitoring detector for spallation neutrons

In a commercial reactor used in the nuclear industry, fission chamber (FC) detectors are used to measure the incore flux profile of thermal neutrons within the reactor core. A fission chamber is a gas-filled ionization chamber to which a fissile coating is added. To make a fission chamber sensitive to thermal neutrons, a coating with the uranium isotope $^{235}$U, which has a fission cross section of 583 barn for thermal neutrons of 0.025 eV, is added in the chamber. On the other hand, $^{239}$Pu fission chambers with a coating of the plutonium isotope $^{239}$Pu, which has a fission cross section of 744 barn for thermal neutrons, are adopted to measure fast neutrons with energy more than 0.1 MeV (e.g. [7]). However, both $^{235}$U and $^{239}$Pu have lower fission cross sections of about 1 barn for fast neutrons. A natural question is whether a fission chamber sensitive to thermal neutrons can be used to measure the fast neutrons from the spallation process?

### 4.1 Virtual fission chamber

An FC consists of a coaxial cable containing an inner electrode (the anode), surrounded by insulation and an outer electrode (the cathode). The inter-electrode space is filled with pressurized gas, such as argon. A thin layer of fissile material is deposited on at least one of the electrodes.

When a neutron reaches the fissile deposit, it is likely to induce a fission that generates two heavily charged ions. The two fission products are emitted in two opposite directions. The one fission product emitted out of the deposit generates a signal by ionizing the internal filling gas. The two operating modes of the FC, which are namely the pulse mode and the Campbell mode, make it possible to perform the flux measurement over a large dynamic range. The Campbell mode is also known as the "fluctuation mode" or the "mean square voltage mode".

In pulse mode, pulse counts generated by the collection of electrons and ions are recorded. The acquisition system consists of a discriminator directly connected to a counter. In Campbell mode, a statistical treatment of the signal is performed. If one only considers the first moment of the Campbell mode, it is the current mode. The acquisition system in Campbell mode consists of a frequency band-pass filter, an analog-to-digital converter and a processing unit for variance calculation.

In the following text, we consider three FCs. The first is a commercial FC, the CFUE32 from PHOTONIS. A fissionable material, uranium with 93%-enriched $^{235}$U, is used for the coating onto the outer surface of the anode, which makes the FC sensitive to thermal neutrons. The FC has a cylindrical geometry with a diameter of 0.7 cm and a sensitive length of 5.6 cm. The inter-electrode gap is filled with argon at 900 kPa. The second FC is an FC in which the uranium with 5%-enriched $^{235}$U and 95%-enriched $^{238}$U is used for the coating onto the outer surface of the anode, which makes the FC sensitive to both thermal and fast neutrons. The third FC is an FC in which natural uranium with 99.27%-enriched $^{238}$U is used for the coating onto the outer surface of the anode, which makes the FC sensitive to fast neutrons. These FCs are referred to as virtual FCs, because only their parameters are used in the following calibration and simulation. The three FCs are named FC-235, FC-mix and FC-238 in the following text.

**4.2 Calibration method for spallation neutron**

FC calibration consists in establishing the relation between the measured indication and the physical quantity. Whatever mode is selected, this relation is linear for FC detectors used in the saturation regime [8]. Several calibration methods for FCs used for thermal neutrons have been reported in the literature (e.g. [8]-[10]). We assume that the FC calibration between the measured indication and the fission rate has been done using a reference neutron source or other method, i.e. we know how to extract fission rate from the measured indication. In this sub-section, we show how to obtain the flux of spallation neutrons from the fission rate.

Letting $n_i$ be the number of nuclei of the isotope $i$ in the deposit, the fission rate $R$ is given by

$$R = \sum_i n_i \bar{\sigma}_i \Phi, \quad (1)$$

where the total neutron flux $\Phi$ and the weighted cross section $\bar{\sigma}_i$ of the isotope $i$ are calculated as follows:

$$\Phi = \int_0^\infty \phi(E) dE, \quad (2)$$

$$\bar{\sigma}_i = \frac{\int_0^\infty \sigma_i(E)\phi(E) dE}{\int_0^\infty \phi(E) dE}. \quad (3)$$

$\phi(E)$ in (2)-(3) is the neutron flux in each energy bin. Note that the weighted cross section depends only on the shape of the energy spectrum, not on the flux. The number of nuclei $n_i$ of the isotope $i$ in (1), i.e. the atomic number density of isotope $i$, is given as follows.

$$n_i = \frac{\gamma_i \rho N_A}{\sum_j \gamma_j M_j}, \quad (4)$$

where $N_A$ is Avogadro's number, $\rho$ is the material density, and $M_j$ is the atom weight of isotope $j$. $\gamma_j$ is the fractional presence (abundance) of isotope $j$. In the fission chamber mentioned in sub-section 4.1, $\gamma_1 = 0.93$ and $\gamma_2 = 0.7$ for U-235 and U-238, respectively.

By re-writing (1), one gets the relation between the fission rate and the flux of spallation neutrons as follows.

$$\Phi = kR, \quad (5)$$

where $k$ is the calibration factor with

$$k = 1 / \sum_i n_i \bar{\sigma}_i. \quad (6)$$

By replacing $\bar{\sigma}_i$ with (3), we re-write (6) as follows.

$$k = \int_0^\infty \phi(E) dE \bigg/ \sum_i n_i \int_0^\infty \sigma_i(E)\phi(E) dE. \quad (7)$$

The calibration method works as follows.

Step 1: Based on the fractional presence of each isotope in the fissile deposit of FC, one calculates the atomic number densities $n_i$ for each isotope according to (4).

Step 2: One obtains the neutron flux $\phi(E)$ in each energy bin, i.e. the energy spectrum, based on the experimental measurement or model calculation.

Step 3: Based on the fission cross sections of each isotope in the fissile deposit of FC, one calculates the calibration factor $k$ for spallation neutrons according to (7).

**4.3 Results for calibration factor $k$**

As an example, first we consider the first FC, FC-235. As described in the calibration method, first one obtains the atomic number densities, $n_1 = 4.54 \times 10^{22}$ and $n_2 = 3.4 \times 10^{21}$ atoms/cm$^3$, for isotopes U-235 and U-238. The neutron flux $\phi(E)$ in each energy bin has been calculated with GEANT4 code in Section 3. Then, based on the fission cross sections of isotopes U-235 and U-238 provided in the ENDF/B-VII data library, we have extracted the calibration factor $k$ of FC for spallation neutrons.

Fig. 4 shows the calibration factor $k$ for FC-235 as a function of the vertical location $z$ of the detector for a 250-MeV proton beam with a current of 10 mA impinging on a tungsten target. The factor $k$ remains unchanged with increasing $z$ value at upper locations from $z = -100$ cm to $z = -10$ cm and at lower locations from $z = 10$ cm to $z = 100$ cm. However, the factor $k$ with negative $z$ value is different from that with positive $z$ value, indicating that different factors are needed for upper and lower locations. For the second and third FCs, FC-mix and FC-238, we have also observed that the factor $k$ remains unchanged with increasing $z$ value at upper locations from $z$ = -150 cm to $z$ = -10 cm and at lower locations from $z$ = 10 cm to $z$ = 150 cm.

Table 1. Calibration factors for three fission chambers.

| Target (cm) | FC-235 | | FC-mix | | FC-238 | |
|---|---|---|---|---|---|---|
| | $k_1$ | $k_2$ | $k_1$ | $k_2$ | $k_1$ | $k_2$ |
| 1 | 15.8 | 15.2 | 60 | 43 | 70 | 48 |
| 3 | 15.2 | 14.8 | 77 | 56 | 95 | 65 |
| 5 | 14.8 | 14.1 | 88 | 67 | 116 | 83 |
| 7 | 14.5 | 13.5 | 93 | 74 | 126 | 94 |
| 10 | 14.4 | 13.0 | 96 | 78 | 132 | 104 |
| 15 | 14.3 | 12.6 | 97 | 79 | 134 | 106 |
| 20 | 14.3 | 12.6 | 97 | 79 | 135 | 106 |
| 30 | 14.3 | 12.6 | 97 | 79 | 135 | 106 |
| 40 | 14.3 | 12.6 | 97 | 79 | 135 | 107 |
| 50 | 14.3 | 12.6 | 97 | 80 | 135 | 108 |
| 70 | 14.3 | 12.6 | 97 | 80 | 135 | 109 |
| 100 | 14.3 | 12.6 | 97 | 80 | 135 | 109 |

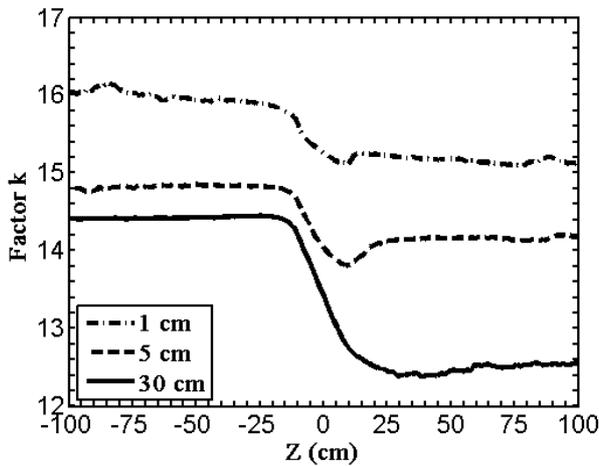

Fig. 4 Calibration factor for FC-235 at different locations $Z$ for 250-MeV proton beam with current of 10 mA impinging on a tungsten target. The results with three target lengths, 1 cm, 5 cm and 30 cm, are compared.

Table 1 lists the averaged factors $k_1$ and $k_2$ of three FCs, where $k_1$ is the factor averaged over all upper locations with $z$<-10cm and $k_2$ is the factor averaged over all lower locations with $z$>10cm. It is shown clearly that the averaged factors keep unchanged with increasing target length from 15 to 100 cm. The calibration factor $k$ at the central location from $z$ = -10 cm to $z$ = 10 cm is taken from the medium value between $k_1$ and $k_2$, i.e. $k = (k_1 + k_2)/2$.

## 5. Analysis in the ADS environment

Since the spallation target is located vertically at the centre of the sub-critical reactor core, the FC for the measurement of spallation neutrons should also be put within the reactor core. Thus, the neutron flux measured by FC comes from two parts, i.e. spallation neutrons from the target and fission neutrons from the reactor. In this section, we assess whether the FC can measure neutron flux well when both spallation neutrons and fission neutrons exist together.

### 5.1 Model for fission neutrons

Prior to the introduction of fission neutrons in the reactor core, we mention the valuable design studies for a FAst Spectrum Transmutation Experimental Facility (FASTEF) by the MYRRHA-FASTEF team [11]. In the FASTEF facility, a proton beam with energy of 600 MeV and beam current of up to 4 mA bombards a Lead-Bismuth Eutectic (LBE) target located at the centre of a sub-critical core cooled with LBE. In the CIADS facility, a proton beam with energy of 250 MeV and beam current of 10 mA bombards a tungsten target located vertically at the centre of a sub-critical core cooled with LBE. Therefore, the fast neutron spectrum in the FASTEF facility is used as the model for fission neutrons in the CIADS facility.

The dots in Fig. 5 represent the fission neutron spectrum within the reactor core, which comes from the simulation results with MCNPX code in [11] (see Fig. 18 in [11]). Since the maximum incore neutron flux is designed for $2 \times 10^{14}$ n/cm$^2$/s for CIADS, the fission neutron flux covers a wide range of 14 decades from 1 n/cm$^2$/s in the startup to $10^{14}$ n/cm$^2$/s in the full-power operation. In order to cover the range of 14 decades, the fission neutron spectrum is normalized to $10^m$ with 1≤$m$≤14 and then is added to the spallation neutron spectra. The fission neutron spectrum in Fig. 5 is normalized to a total flux of $10^{13}$ n/cm$^2$/s. The solid and dashed lines in Fig. 5 represent the total flux of both spallation neutrons and fission neutrons at vertical location $z$ = -30 and 0 cm, respectively, when fission neutrons with a flux of $10^{13}$ n/cm$^2$/s is added.

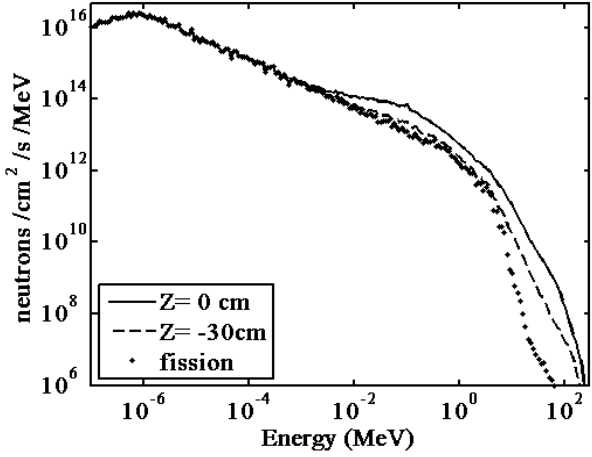

Fig. 5. Neutron energy spectrum for a tungsten target with a length of 30 cm. The dots represent the simulation results with MCNPX code for fission neutron spectrum within the reactor core [11]. The fission neutron spectrum is normalized to a total flux of $10^{13}$ n/cm²/s. The solid and dashed lines represent the total flux of both spallation and fission neutrons at vertical location $z$ = -30 and 0 cm, respectively, where the spallation neutrons are based on the Geant4 simulation.

### 5.2 Simulation results

Fig. 6 shows the total neutron flux as a function of fission neutron flux. The solid lines represent the total flux used in the model described in Sub-section 5.1, where the spallation neutrons come from the GEANT4 simulation for a tungsten target with a length of 30 cm bombarded by 250-MeV protons with a current of 10 mA. The open circles, stars and open squares represent the total flux measured by the three FCs, respectively. The so-called "measured" flux is obtained according to (5), where $k$ is the calibration factor listed in Sub-section 4.3 and $R$ is the fission rate "measured" by the three virtual FCs. The "measured" fission rate $R$ has been estimated as follows.

$$R = \int_0^\infty [n_1\sigma_1(E) + n_2\sigma_2(E)]\phi(E)dE , \qquad (8)$$

where $n_1$ and $n_2$ are the numbers of isotopes U-235 and U-238 in the fission chamber, respectively; while $\sigma_1$ and $\sigma_2$ are the fission cross section of U-235 and U-238, respectively. $\phi(E)$ in (8) is the total flux of both spallation neutrons and fission neutrons in each energy bin.

It is shown clearly in Fig. 6 that the three vitual FCs with the proposed calibration method can measure well the total neutron flux, when the fission neutrons have a flux of less than $10^{13}$ n/cm²/s. This is because the fraction of thermal neutrons is very small, in comparison with the spallation neutrons with a flux of $10^{13}$ n/cm²/s. Although the uranium isotope $^{235}$U has a fission cross section of 583 barn for the thermal neutrons, the fission events occurring in the FC mainly come from the spallation neutrons. In other words, the calibration factor extracted from the spallation neutron spectra can be used well at startup and during low power operation.

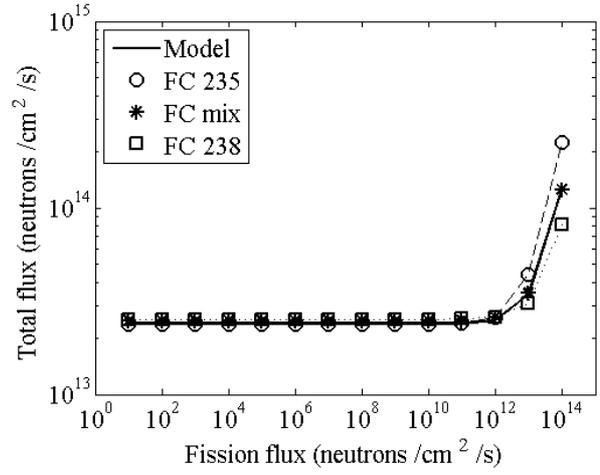

Fig. 6. Neutron flux spectrum at the central location for a tungsten target with a length of 30 cm. The total flux includes both spallation neutrons and fission neutrons.

When the fission neutrons have a flux of more than $10^{13}$ n/cm²/s, i.e. when the reactor operates at full power, the fraction of thermal neutrons is not small, in comparison with the fast neutrons or spallation neutrons. The neutrons into the FCs come from both thermal neutrons and fast neutrons. The results with the open circles indicate that the FC coated by the uranium with 93%-enriched $^{235}$U overestimates neutron flux, because the uranium isotope $^{235}$U within the FC has a much higher fission cross section for thermal neutrons than fast neutrons. To use such a commercial fission chamber in the ADS system, a new calibration factor extracted from the total neutron spectra should be used when the reactor operates at full power.

On the other hand, the results with the open squares indicate that the FC coated by the uranium with 99.27%-enriched $^{238}$U underestimates neutron flux, because the uranium isotope $^{238}$U has a very low fission cross section of about $10^{-4}$ barn for neutrons of less than 1 MeV. It is difficult for the FC coated with $^{238}$U to detect these neutrons of less than 1 MeV. Finally, if the uranium with 5%-enriched $^{235}$U and 95%-enriched $^{238}$U is used for the coating, the results with stars in Fig. 6 indicate that the FC can estimate neutron flux well.

## 6. CONCLUSIONS

We have investigated a monitoring method for neutron flux for the spallation target used in an ADS system. Two main contributions are summarized as follows. For the monitoring method, we have proposed a multi-point measurement technique, i.e. the neutron flux should be measured at multiple vertical locations, because of the difference of neutron flux at difference locations. The detectors for the measurement of spallation neutron flux may either be left in a fixed location or provided with a motorized drive to allow vertical movement within the reactor core. By performing simulations with GEANT4 codes for a tungsten target bombarded by a 250-MeV proton beam we have observed that the difference between neutron flux at the central location and at the lower locations is up to three orders of magnitude. This is because the central location is much closer to the target than the lower or upper locations. Furthermore, due to neutron absorption by the target, the neutron flux at the lower locations decreases dramatically with increasing target length.

For the monitoring detector, we have proposed an effective calibration technique for the fission chamber (FC) in which a thin layer of fissile material is deposited on an electrode. Three coating materials, $^{235}$U, $^{238}$U and uranium with 5%-enriched $^{235}$U and 95%-enriched $^{238}$U, have been compared. In order to accurately measure the spallation neutron flux with a FC, we have developed an effective calibration technique by establishing the relation between the fission rate measured by FC and the spallation neutron flux. By evaluating the proposed calibration method for FC in the environment of an ADS system, we have observed that a FC with a coating material of 5%-enriched $^{235}$U and 95%-enriched $^{238}$U can measure spallation neutron flux well.

## REFERENCES

[1] S. Agostinelli et al., Geant4 – a simulation toolkit, Nucl. Instr. Meth. Phys. Res. A 506 (3) (2003) 250–303, http://dx.doi.org/10.1016/S0168-9002(03)01368-8.

[2] J. Allison et al., Geant4 developments and applications, IEEE Trans. Nucl. Sci. 53 (1, Part 2) (2006) 270–278, http://dx.doi.org/10.1109/TNS.2006.869826.

[3] Y. Malyshkin, I. Pshenichnov, I. Mishustin, T. Hughes, O. Heid, W. Greiner, "Neutron production and energy deposition in fissile spallation targets studied with Geant4 toolkit", *Nucl. Instr. Meth. Phys. Res.* B289, pp. 79-90, 2012.

[4] Y. Malyshkin, I. Pshenichnov, I. Mishustin, W. Greiner, "Monte Carlo modeling of spallation targets containing uranium and americium", *Nucl. Instr. Meth. Phys. Res.* B334, pp. 8–17, 2014.

[5] G. Forger, V. N. Ivanchenko, and J. P. Wellish, The binary cascade. The European Physical Journal A. 21, pp. 407-417, 2004.

[6] M.M. Meier, C.A. Goulding, G.L. Morgan and J.L. Ullmann, Neutron yields from stopping- and near-stopping-length targets for 256-MeV protons, Nucl. Sci. Eng. 104 (1990), pp. 339-363.

[7] ZENG Li-Na, WANG Qiang, SONG Ling-Li, ZHENG Chun., Chin. Phys. C, 2015, 39(1), 016001.

[8] S. Chabod, "Saturation current of miniaturized fission chambers," Nucl. Instrum. Methods Phys. Res., vol. 598, no. 2, pp. 577–590, Jan. 2009.

[9] Benoit Geslot, Troy C. Unruh, Philippe Filliatre, Christian Jammes, Jacques Di Salvo, Stéphane Bréaud, and Jean-François Villard, "Method to Calibrate Fission Chambers in Campbelling Mode", *IEEE Trans. on Nuclear Science*, vol. 59, no. 4, pp. 1377-1381, 2012.

[10] Zs. Elter, C. Jammes, I. Pázsit, L. Pál, P. Filliatre, "Performance investigation of the pulse and Campbelling modes of a fission chamber using a Poisson pulse train simulation code, " *Nuclear Instruments and Methods* A 774, pp. 60-67, 2015.

[11] Massimo Sarotto, et. Al., "The MYRRHA-FASTEF cores design for critical and sub-critical operational modes," *Nuclear Engineering and Design*, vol. 265, pp.184-200, 2013.